\documentclass[prl,floatfix,reprint]{revtex4}
\usepackage{graphicx,amssymb,amsmath}
\usepackage{booktabs}
\newcommand{\ra}[1]{\renewcommand{\arraystretch}{#1}}

\begin{document}

\title{Understanding the variability of daily travel-time expenditures using GPS trajectory data}

\author{Riccardo Gallotti\footnote{Correspondance to: rgallotti@gmail.com}}
\affiliation{Institut de Physique Th\'{e}orique, CEA-Saclay, Gif-sur-Yvette, France}
\author{Armando Bazzani}
\affiliation{Department of Physics and Astronomy, University of Bologna, INFN Bologna section}
\author{Sandro Rambaldi}
\affiliation{Department of Physics and Astronomy, University of Bologna, INFN Bologna section}

\begin{abstract} 

Transportation planning is strongly influenced by the assumption that every individual has
for his daily mobility a constant daily budget of $\approx$1 hour. However, recent experimental results are proving this assumption as wrong.
Here, we study the differences in daily travel-time expenditures among 24
Italian cities, extracted from a large set of GPS data on vehicles
mobility. To understand these variations at the level of individual behaviour, we
introduce a trip duration model that allows for a description of
the distribution of travel-time expenditures in a given city using
two parameters. The first parameter reflects the accessibility of
desired destinations, whereas the second one can be associated to a travel-time budget and represents
physiological limits due to stress and fatigue. Within the same
city, we observe variations in the distributions according to home
position, number of mobility days and a driver's average number of
daily trips. These results can be interpreted by a stochastic
time-consumption model, where the generalised cost of travel times
is given by a logarithmic-like function, in agreement with the
Weber-Fechner law. Our experimental results show a significant variability in the travel-time budgets in
different cities and for different categories of drivers within the same city. This explicitly clashes with the idea
of the existence of a constant travel-time budget and opens new perspectives for the modeling and governance of urban mobility.

\end{abstract}

\maketitle

\section*{Introduction}

Recently, human mobility has been extensively studied using data on
individual trips provided by the information-communication
technologies~\cite{gonzalez2008,song2010,cheng2011,noulas2012,vespignani2012,hawelka2014,lenormand2014}.
In mobility-related decisions, travel time appears as a natural cost
function, since it represents a limited resource used for performing
daily activities~\cite{axhausen2008}. The concepts of Travel-Time
Expenditure (TTE, the daily amount of time spent traveling) and
Travel-Time Budget (TTB, the average daily amount of time that
people make available for mobility~\cite{zahavi1974}) have been
introduced by transportation planners to model the mobility demand
and to explain some of the features characterising urban
mobility~\cite{stopher2011}. Travel-Time Expenditure and Budget are
more comprehensive quantities than the commuting time from home to
work and back between home and work, and the related concept of
Marchetti's constant~\cite{marchetti1994,kung2014}. Indeed, this second
perspective is limited to the journey-to-work mobility and thus
exclude a large fraction of the individuals' mobility demand
associated to amenities.

The existence of a Travel-Time Budget is assumed on the basis of the behavioural
hypothesis that people spend a fixed amount of time available on
traveling~\cite{mokhtarian2004}. The extreme interpretation of Travel-Time Budget
as an universal constant stable in space and time is still sustained
and very influential in urban planning. Indeed, if Travel-Time Budget is constant, any investments in better infrastructure would not reduce daily travel times (and possibly, through that, polluting emissions) since it would only create new induced travel demand~\cite{cervero2011}.
Most of the
empirical results on Travel-Time Budget are determined as average values from large
travel surveys. At a disaggregate level, however, Travel-Time
Expenditures appear
strongly related to the heterogeneity of the individuals, to the
characteristics of the activities at destinations and to the
residential areas~\cite{mokhtarian2004}. Aggregated results suggest
that the average amount of time spent traveling is constant both
across populations and over time: approximatively 1.0-1.1 hours per
day~\cite{metz2008}. Despite the gains in average travel speed due
to infrastructural and technological advances in the past decades,
Travel-Time
Expenditures appear more or less stable or even
growing~\cite{tooleholt2005,vanwee2006,milthorpe2007}. This growth
can be associated to the super-linear relationship between a city's
population and the delays due to congestion~\cite{Louf:2014}.

In Italy, Global Positioning System (GPS) devices are installed
in a significative sample of private vehicles for insurance reasons.
The initial and the final points of each trajectory are recorded, together with the path length and some intermediate points at a
spatial distance of 2 km or at a time distance of 30 seconds. These
data allow a detailed reconstruction of individual mobility in
different urban contexts~\cite{gallotti2012} and measure the elapsed
of time during mobility~\cite{gallotti2015}.

In this paper, we explore the statistical
features of Travel-Time
Expenditures related to private mobility, both from an aggregate
and individual point of view. Our goal is to point out some of the
factors influencing travel demand by means of new specific measures,
which describe differences among cities. The statistical analysis of
empirical data points to the existence of a universal law underlying
the distributions of Travel-Time
Expenditures, which highlights the nature of time
constraints in vehicular mobility.
This result allows us to observe in detail the differences in daily travel demand for different cities, challenging
the idea of a constant Travel-Time Budget and pointing out the important role of accessibility~\cite{hansen1959}.

\begin{table*}[h!]
\begin{tabular*}{1\textwidth}{@{\extracolsep{\fill}}lcccccccccccc@{}}
\toprule[1pt]
Quantity  & Notation & Abbreviation\\
\hline
Daily travel-time expenditure &  $T$ & TTE\\
Daily travel-time budget  & $\beta$ & TTB\\
Accessibility time & $\alpha$ & - \\
Single trip travel-time & $t$ & - \\
\toprule[1pt]
Function  & Notation & Abbreviation\\
\hline
Probability density of $x$ & $p(x)$ & PDF\\
Cumulative density of $x$ ($\int^x p(x') dx'$)& $P(x)$ & CDF\\
Survival function ($1-P(x)$) & S(x) & - \\
Hazard function ($dS(x)/dx$)& $\lambda(x) $ & -\\
Conditional probability of $x$ given $y$ & $\pi(x|y)$ & -\\
\bottomrule[1pt]
\end{tabular*}
\caption{List of notations.}
\end{table*}

\section*{Assumptions}

\begin{figure*}[h!]
\includegraphics[angle=0, width=0.9\textwidth]{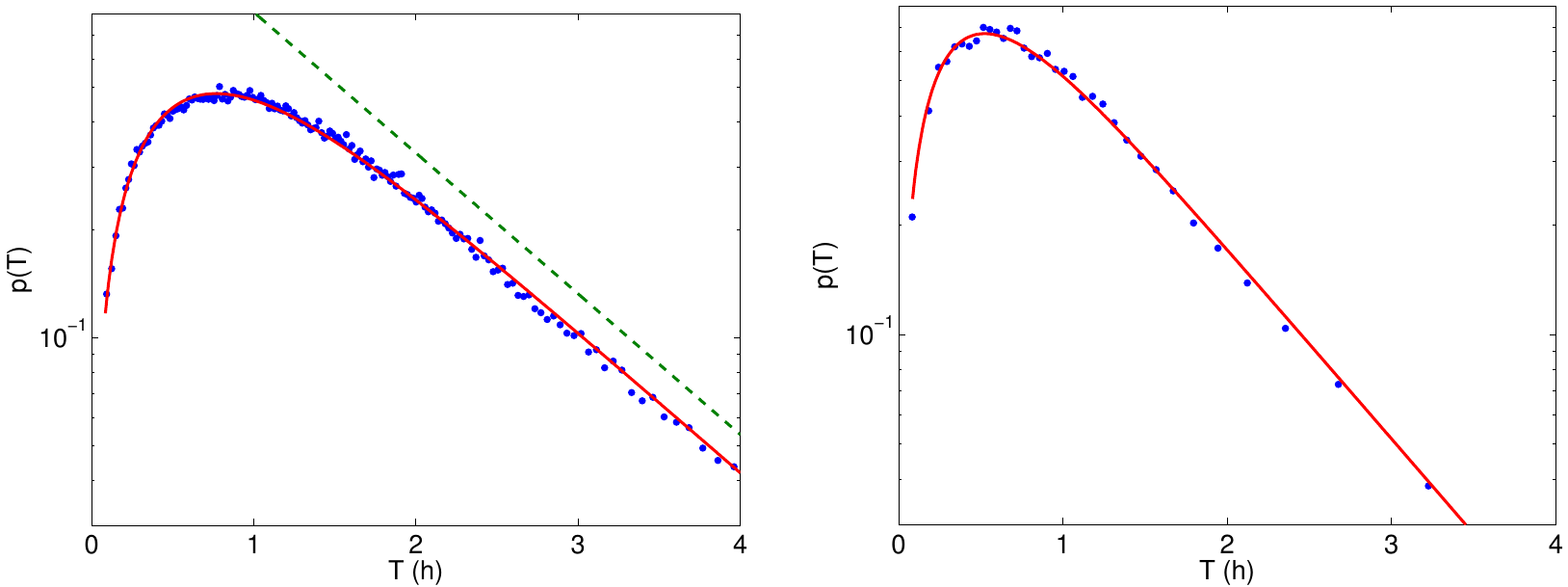}
\caption{ \textbf{(Left) Travel Time Expenditure
(TTE) distribution for the city of Naples ($\approx 1$ million
inhabitants)}. The empirical probability density $p(T)$ (dots) is
correctly interpolated by the curve (\ref{analyticaldistribution})
(solid line) with $\alpha = 0.61$h, $\beta = 1.11$h and $R^2 =
0.99$. The dashed line, shown as a guide to the eye, represents an
exponential decay with a characteristic time $\beta$.
\textbf{(Right) TTE distribution for the city of Grosseto ($\approx
80,000$ inhabitants)}. The interpolation with the curve
(\ref{analyticaldistribution}) is successful also for smaller cities
like Grosseto: in this case the parameter values are $\alpha =
0.38$h, $\beta = 0.83$h and $R^2 = 0.99$ } \label{fitdistribution}
\end{figure*}

Previous empirical observations on different
data-sources~\cite{kolbl2003,bazzani2010,gallotti2012,liang2012,liang2012b}
have shown out that the TTE probability distribution $p(T)$, associated
to a single mean of transportation, is characterized by an
exponential tail
\begin{equation}
p(T) = \beta^{-1}\exp(-T/\beta)\,, \qquad \qquad \textrm{for} \ T
> 1\;\textrm{hour.}
\label{expdist}
\end{equation}
where $\beta$ is a fit parameter. Our analysis confirms the
universal character of the exponential behaviours for the TTE
empirical distribution and points out relevant differences among the
considered cities (see Fig.~\ref{fitdistribution}).

As it is well known from Statistical Mechanics,
the exponential distribution (\ref{expdist}) can be derived from the
Maximal Entropy Principle under some minimal
assumptions~\cite{gallotti2012}. More precisely, one assumes the
existence of an average finite TTE for the considered population and
the independence of the individual behaviour: i.e. any microscopic
configuration which associates a TTE to each individual with the
constraint that the average TTE is finite, has the same probability
to be observed. The parameter $\beta$ defines the average time scale
that limits the individual TTE and we will show that this is a characteristic of each
city. Therefore, we propose to associate the concept of TTB to the
value $\beta$ which characterizes the exponential decay of the daily
travel-time distribution. However, the eq.~(\ref{expdist}) does not give
information on the dynamical processes underlying the human mobility
which produces the distribution. We take advantage from the
dynamical structure of the GPS data to propose a duration model (see Methods) that
seems to be endowed with universal features with respect to the
considered cities. The essential hypotheses at the bases
of the duration model are: i) it exists a TTB; ii) the
individual decision to continue the mobility for a time $\Delta T$
after a TTE $T$ is the realization of a independent random event
whose probability decrease proportionally to $\Delta T$.

\section*{Results}

\subsection*{The variability of Travel Time Expenditures}

The average value of TTE does not give a
sufficient insight on the statistical features of the distribution
$p(T)$.
For each
city the statistical features of the distribution $p(T)$ turn out
to be characterized by the two time scales $\alpha$ and $\beta$.
In the duration model, after a characteristic time  $\alpha$, the choice of going back home
or proceeding with further extra traveling is limited by the
available TTB, whose average value is quantified by the time scale
$\beta$. $\alpha$ therefore represents the average time under which the use of a
private car seems to be not convenient. 
 
\begin{figure*}[h!]
\begin{center}
\includegraphics[angle=0, width=0.9\textwidth]{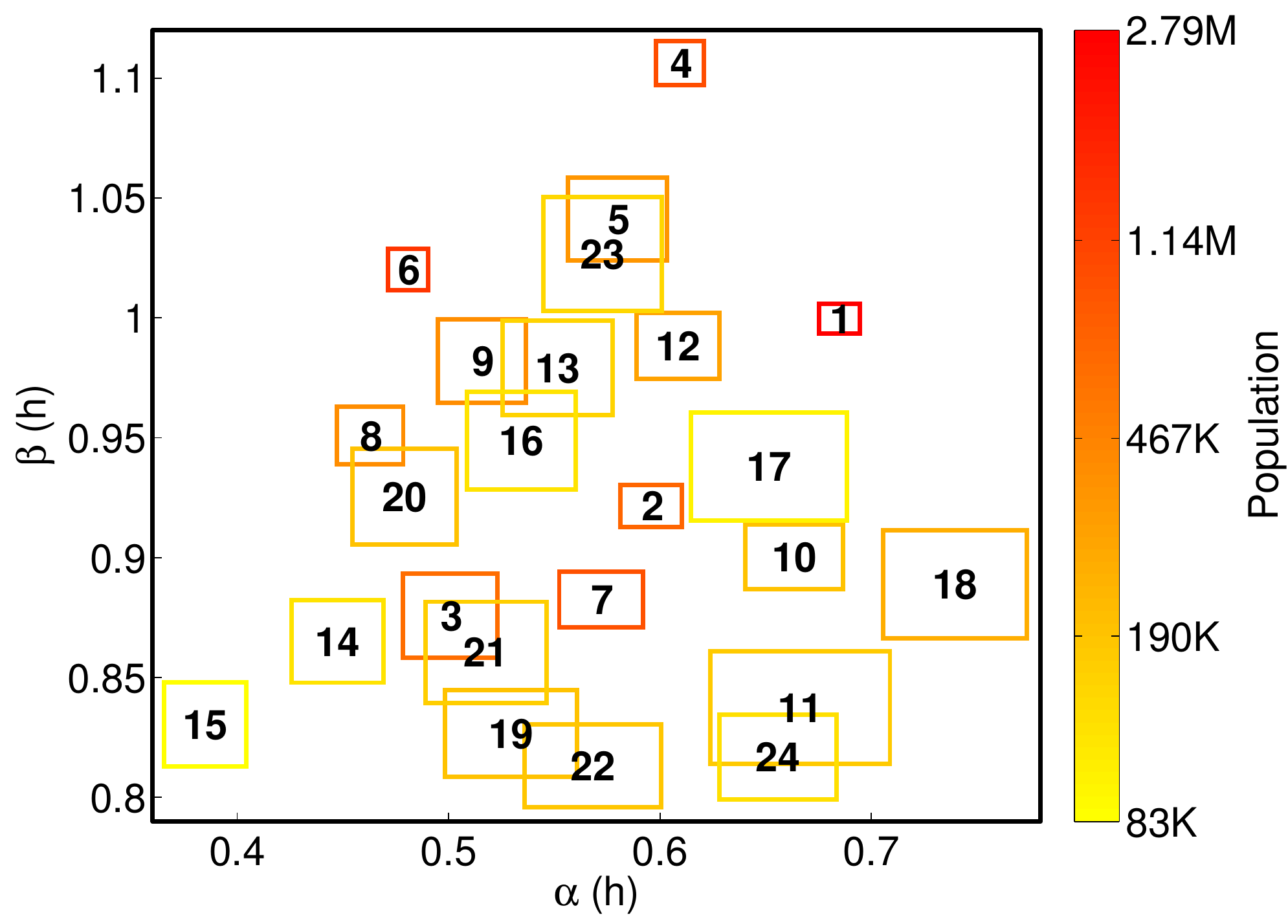}
\end{center}
\caption{\textbf{Values of $\alpha$ and $\beta$ for the 24 cities studied.} The boxes represent 95\% confidence intervals obtained with a bootstrap, for empirical best-fits for the model parameters. The differences we observe in the accessibility time $\alpha$ and in TTB $\beta$ are thus significant and uncorrelated ($r$ = 0.09). Both timescales are weakly correlated with the city's population ($r$ = 0.20 for $\alpha$ and $r$ = 0.40 for $\beta$.) }
\label{scatter}
\end{figure*}T

We focus our study on 24 Italian cities where we had a large statistic of
users. The values of $\alpha$ and $\beta$ are estimated from a bestfit for S(T) with equation S5 (which is equivalent to fitting the CDF). The results are displayed in Fig.~\ref{scatter} and reported in the Table S1. Two examples are also proposed in Fig.~\ref{fitdistribution}.
The two parameters are independent, with a Pearson correlation
coefficient $r = 0.09$. $\beta$ fall in the interval 0.8-1.1h, which
is reasonably consistent with the values reported in the
literature~\cite{metz2008}. Nevertheless, the differences we observe
among cities are statistically significant, as the 95\% confidence
intervals for the fits, estimated with a bootstrap, are $\leq0.02$h. 
Therefore, our results clearly clash with the concept a constant
TTB. 

Since the values of $\beta$ are moderately correlated with the number of inhabitants of the municipality ($r$ =
0.40) or population density ($r$ = 0.49), some of
this variability is dependent on the city
population~\cite{bettencourt2007,Louf:2014}. 
The accessibility time
$\alpha$ is only weakly correlated with city population ($r$ =
0.20) and not correlated ($r$ =
0.03) with population density, and fall in the interval 0.3-0.8h. 
The confidence intervals
for the fits are $\leq0.04$h, granting that we have significant
differences in accessibility time among cities. The general picture
displayed in Fig.~\ref{scatter} shows that, if one has appropriate
datasources to characterize the daily mobility of a single city, one
needs the knowledge of both parameters. Under this lens, the
variability of TTE is manifest and can be observed in both the
ramping part ($\alpha$) and the tail ($\beta$) of the distribution.

\subsection*{Disaggregate analysis: The case of Milan}

\begin{figure*}
\includegraphics[angle=0, width=0.9\textwidth]{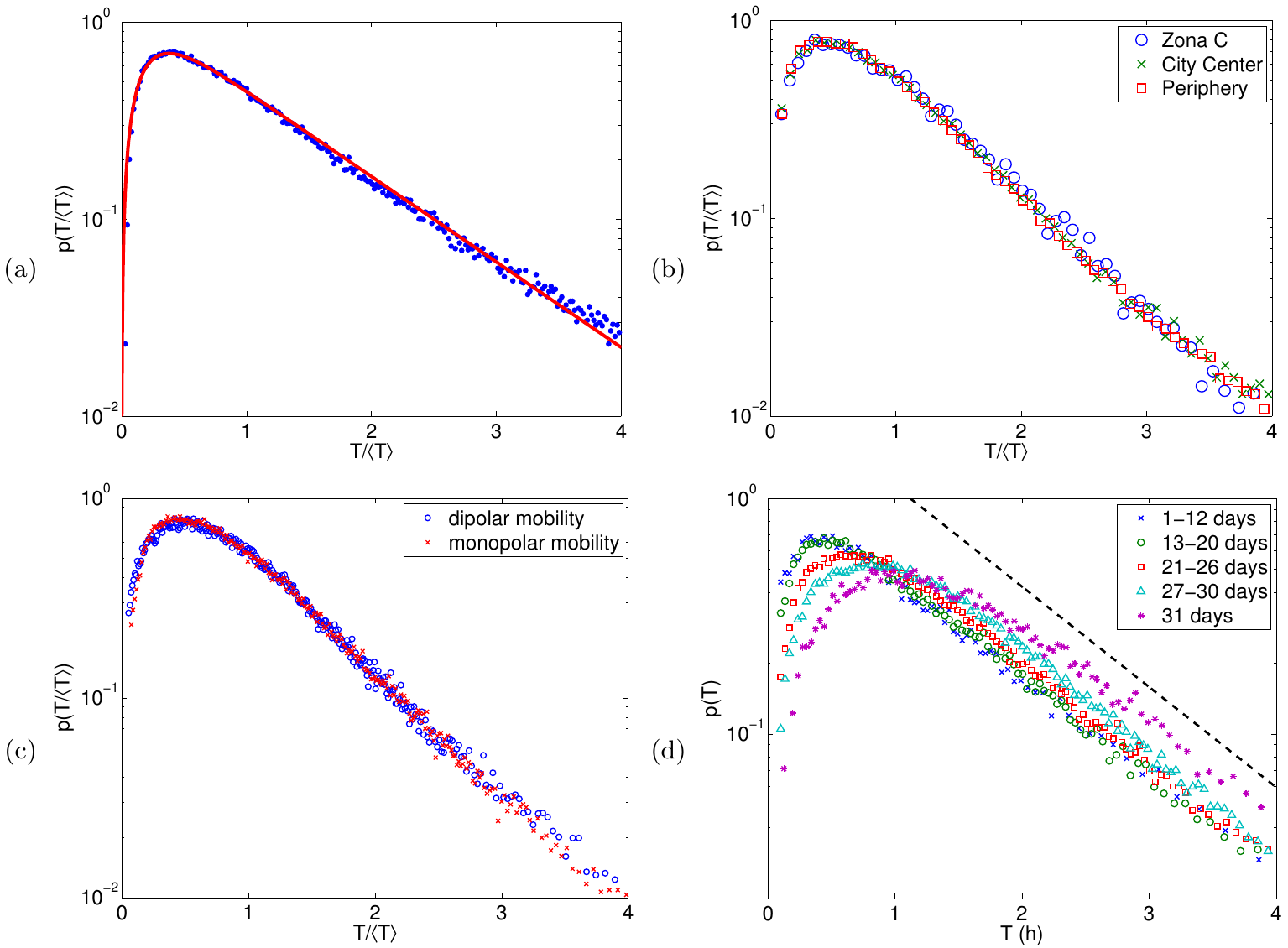}
\caption{ \textbf{a) Distribution of the normalised individual TTE
for the city of Milan}: the distribution of normalised individual
TTE can be represented by the analytical distribution
(\ref{analyticaldistribution}). The solid line is obtained by using
the parameters $\alpha = 0.19$ and $\beta = 1.00$. \textbf{b)
Normalised TTE distributions disaggregated according to home
location}: the position of the main mobility hub (home) influences
the average value of $T$ but not the distribution scaled by that
value. For the 'Zona C' we have $\langle T \rangle =
1.520\pm0.009$h, for the city center we have $\langle T \rangle =
1.482\pm0.004$h while for the periphery we have $\langle T \rangle =
1.416\pm0.003$h (errors correspond to the s.e.m); \textbf{c)
Normalised TTE distributions disaggregated according to
classification of the mobility network}: the role of home influences
only the average value of $T$: selecting people whose mobility is
characterised by more than one hub in their mobility network (we
look for people with a percentage greater than 25\% of round trips
not starting and ending at home). Such people have $\langle T
\rangle = 1.72$h, whereas people with a single mobility hub have
only $\langle T \rangle = 1.24$h. \textbf{d) TTE distributions
disaggregated according to the number of mobility days}: these
disaggregated distributions have a similar decaying in the tails,
i.e they have similar value of $\beta$ (as a guide for the eye, the
dashed line represents an exponential decay with the characteristic
timescale $\beta$ for Milan). Differences emerge instead in the
behaviour of short TTE: people that use the car more regularly, have
longer TTEs, since they perform more trips.}
\label{individualgraphs}
\end{figure*}

Macroscopic statistical laws might depend on the details of the
microscopic dynamics. Their extension down to the interpretation of
the individual behaviour is therefore under
debate~\cite{zhou2013}. 
Nevertheless, we believe that the universal character inherent to the
concept of TTB could
be an individual property. To support this statement, we consider
here a disaggregate analysis of the GPS mobility data suggesting that
 our results might be extended to the individual level. A
limitation of this analysis comes from the limited time considered in our dataset.  Indeed, it refers
only to a single month of mobility, a period probably too short to infer a
definitive conclusion on our hypothesis.

We study the case of the city of Milan, the largest city in North
Italy with $\approx 1.3$ millions inhabitants (dash line in
Fig.~\ref{differentTTE} left and labeled 6 in Fig.~\ref{scatter}).
We start with verifying that the shape of TTE distribution $p(T)$ is
a property of each single individual. Using the GPS data, the
heterogeneity of the population can be quantified by considering the
distribution of the average individual TTE $\langle T \rangle$
empirically computed from the individual daily mobility. To compare
different individuals, we normalize each TTE value by the
corresponding individual average. In Fig. ~\ref{individualgraphs}
(a), we show that the distribution of the normalised individual TTE
$p(T/\langle T \rangle)$ is still very well fitted by the analytical
curve (\ref{analyticaldistribution}). Therefore we conjecture that
$\langle T \rangle$ contains the relevant information to explain the
individual heterogeneity and the distribution
(\ref{analyticaldistribution}) has an universal character that
extends up to the individual level.

An individual disaggregation, according to the home location or
characteristics of the mobility network, confirms the previous
hypothesis (see Figs.~\ref{individualgraphs} (b) and (c), and the
Supplementary Information for further details). Thus, the
heterogeneity is mainly determined by the average value $\langle T
\rangle$ evaluated within each class. However, we find that $\langle
T \rangle$ is longer for:

i) people living in the city center ($\approx 8\%$ longer than for
people living in the periphery), a result consistent with what was
found in Ref.~\cite{milthorpe2010} for the city of Sydney;
conversely, people in the periphery tend to make $\approx 0.5$ trips
more per day and $\approx3.3$ days more of mobility in average;\par
ii) people performing many round trips (A-B-A patterns) not
involving home.
\par\noindent
The last criterium points to the existence of a second center of
daily activity and allows to separate individual mobility networks
into mono-centric and polycentric ones~\cite{gallotti2013}.
\par\noindent
Our empirical data suggest that people with a polycentric mobility
(who have more than one mobility hub) have greater $\langle T
\rangle$ than people whose round trips start and end at home.
However, if we classify the individuals according to the number of
days in which they used the car, the TTE distributions differ when
we consider small $T$ values (see Fig.~\ref{individualgraphs} (d)).
Even if the exponential tail of the distributions does not
change significantly, a tendency to suppress more short values of
$T$ is observed for users accustomed to regularly carry out their
daily mobility by car. Our duration model associates this to a
larger value of $\alpha$ and therefore the need in average of longer
times to accomplish the necessary tasks of the day. In summary,
people who take the car more often also need to drive more, yet
maintaining a similar TTB. This is confirmed by considering the
number of trips $n$ that are accomplished in a day. The average
value of $n$ grows from $4.2$, for people who drove 1-12 days up to
the $7$ for the class of users who drove all the 31 days (see
Supplementary Information). This result clearly links the value of
the accessibility time $\alpha$ to the need of accessing to the
desired destinations by car. Drivers who experience better
accessibility do not need to use the car every day, and when they do
they can also drive less.  In the following, we show that  these
differences can be linked to a different value of time for users
performing more trips.

\subsection*{Evidence of a log-perception of travel-time costs}

\begin{figure*}
\centerline{
\includegraphics[angle=0, width=0.9\textwidth]{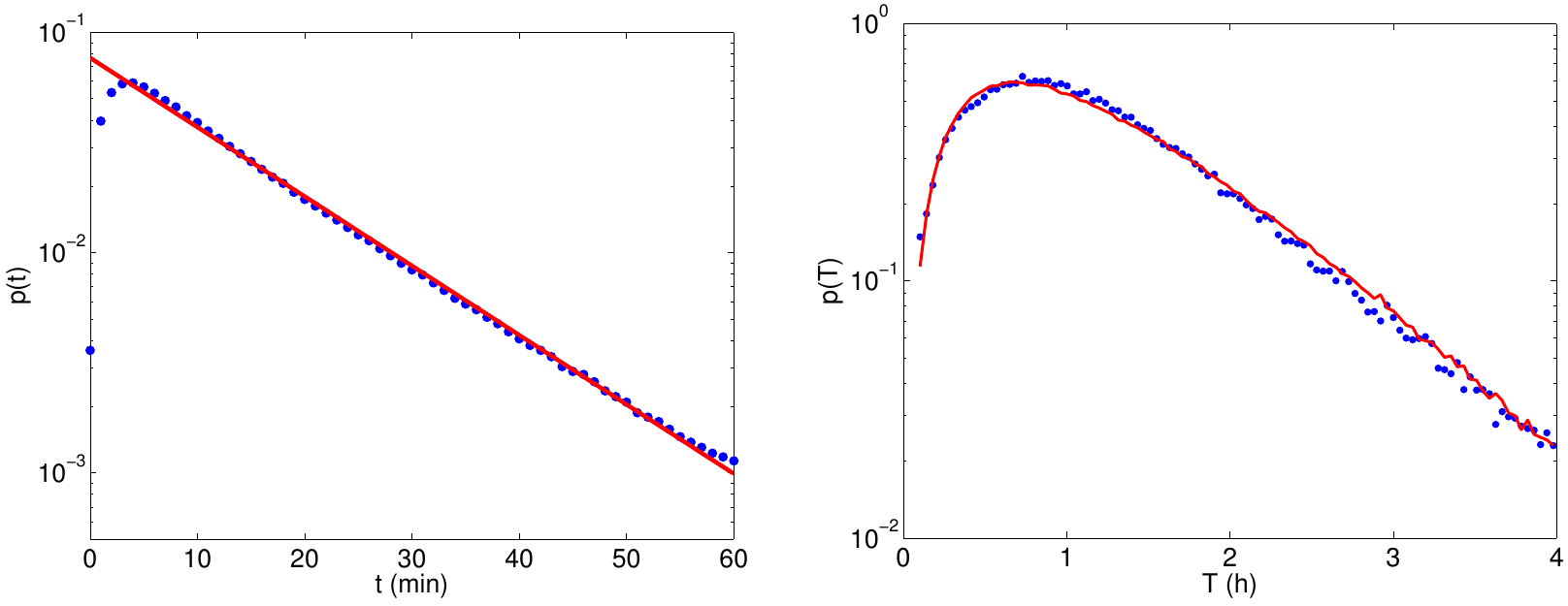}
} \caption{\textbf{Travel-time distribution and stochastic decision model.}
{\bf (Left)} The travel-times distribution $p(t)$ in Milan (dots)
compared with an exponential interpolation (solid line). The
suppression for short travel-times, with $t < 4$ min could be a
consequence of the characteristic GPS measurement time and does not
affect the time scale $\langle t\rangle$. The results are consistent
with the exponential fit of the tail. {\bf (Right)} Comparison
between the empirical TTE $p(T)$ distribution in Turin (dots) and
the best fit distribution provided by our stochastic decision model,
using a logistic threshold function (solid line, see Supplementary
Information, $R^2 = 0.99$).} \label{numerical}
\end{figure*}

Finally, to link the duration model with the individual behavior and  to shed light on the how
individuals organize their mobility, we
propose a time consumption model that allows for an interpretation of
the empirical observations compatible with a logarithmic perception
of the time cost of a trip. (See Methods). This model realizes a
stochastic process for the individual decisions at the base of TTE,
which is compatible with the assumptions of the duration model and
with a possible logarithmic perception of the cost $\Delta T$ of a
trip, analogous to the Weber-Fechner psychophysical
law~\cite{dehaene2003}. The same model does not
reproduce the empirical observations, assuming a linear time
perception.  This result has been confirmed with a Montecarlo stochastic decision model based on the same premises (see Fig.~\ref{numerical} left and Supplementary Information). 
This model assumes a logit curve~\cite{domencich1975} in the decision model of the binary choice of interrupting the daily mobility after a certain trip, we could fit the TTE distributions in all cities with great precision ($R^2>0.986$). 

The existence of simple universal dynamical
models for empirical TTE distribution allows to introduce few
observables that point out relevant differences among cities and
suggests relations between the presence of mobility infrastructures
and/or the socio-economic indexes of a city, and the features of the
empirical TTE distribution. 
These relations could be useful for urban planners to build governance policies for mobility.

\section*{Discussion}

In our analysis, based on a large GPS database containing
information on single vehicle trajectories in the entire Italian
territory, we point out that the empirical distributions for the
daily Travel-Time Expenditures in different cities can be modeled by a single distribution. This distribution is function
of two parameters: $\alpha$ and $\beta$. The
time scale $\alpha$ measures the characteristic mobility time
associated to the use of private cars in a given city, whereas the
limit value $1/\beta$ of the hazard function $\lambda(T)$ as $T\gg
1$, is associated to the concept of Travel-Time Budget. In our
opinion, $\alpha$ is a good measure of the average
accessibility~\cite{hansen1959} of a city. Lower values of $\alpha$
(i.e. higher accessibility) mean a better proximity to useful
locations and faster mobility. We
remark that if one considers Italian cities of different size and
socio-economical conditions, the shape of the distribution appears
to be endowed with an universal character only the values of
Travel-Time Budget and suppression of short Expenditures changes.

The distribution $p(T/\langle T\rangle)$ has an universal character. This suggests the existence of a behavioural model for the
urban mobility that mimics the individual decision mechanisms.
As a consequence, the statistical properties pointed out by the distribution
(\ref{analyticaldistribution}) are traits of the individual
behaviour
and the aggregated probability distribution for a city is averaging over the individual heterogeneity in the values of $\alpha$ and $\beta$ across the population.
However, in the disaggregated analysis of GPS data at individual level, we find
significant differences in the average Travel-Time Expenditure for
different categories of drivers.  In particular, drivers which use their car
more often have higher values of $\alpha$ even if their $\beta$ is
approximatively the same (see Fig.~\ref{individualgraphs} (d)). This
is another confirmation of our interpretation of the parameter
$\alpha$ as a measure of accessibility, because who has the worst
accessibility to public transport facilities is forced to use the
private vehicle over wider range of travel-times.

To interpret these results we propose a simple decisional model, which
assumes the the existence of a \textit{mobility energy} (the
daily travel-time) and a log-time perception of
the travel-time cost for a single trip. These results are also consistent with the Benford's
empirical distribution of elapsed time during human
activities~\cite{gallotti2012} and Weber-Fechner psychophysical
law~\cite{takahashi2008}. Using a Statistical Mechanics point of view, the Travel-Time
Expenditure $T$ plays the role of energy in a model of the
individual urban mobility based on a generalised utility function.
However, one cannot simply define the trip duration $\Delta T$ as a
mobility cost, because the data suggest that this perceived cost
seems to decreases as the daily travel time $T$ grows. A time
consumption model that assumes a scaling cost $\propto \Delta T/T$
(i.e. a {\it law of relative effect}~\cite{helbing2010})
corresponding to a logarithmic preference scale~\cite{dehaene2003}),
is able to reproduce the statistical properties of the empirical
observations.  As a direct application of this result, we are able
to suggest the use of a non linear value of time for the
activity-based modeling of human mobility.

At city-aggregate level, we observe that for every city the average
Travel-Time Expenditure $\langle T \rangle$ is greater than the
Travel-Time Budget $\beta$, because short values of $T$ are
statistically under-expressed~\cite{kolbl2003}. This could reflect
both the fact that the individual mobility demand is hardly
satisfied after short travel-times, and the disadvantage using a
private car for short times. Both $\alpha$ and $\beta$ are
needed to fully understand the Travel-Time Expenditures in a city.
A straightforward application of the approach we propose permits to highlight
the differences in the travel-time expenditures among cities and classes of individuals.
In particular, we clearly observe a variability in the Travel-Time
Budget $\beta$ among cities. The dependency upon population density
and the differences observed in the disaggregate analysis explicitly
clashes with the idea of the existence of a fixed Travel-Time Budget.

Our results intend to nourish the discussion against this old paradigm of a constant Travel-Time Budget, which dangerously
suggests that is not possible to reduce travel times, and therefore
CO2 emissions, with improvements to the transportation
infrastructures. The idea that travel time savings are not beneficial, because
improving road infrastructures in cities will attract even more
traffic, is not corroborated by the empirical data.
Understanding the decision mechanisms underlying the individual
mobility demand and the use of private vehicles in a city is a
fundamental task to forecast the impact of new transportation
infrastructures or of traffic restriction policies. On our opinion,
it is thus clear the need of replacing constant the travel time
budget and induced travel demand assumptions with new models, which
should necessarily encompass both individual behaviour and city
development.

\section*{Methods}

\subsection*{GPS database}
\label{data}

This work is based on the analysis of a large database of GPS
measures sampling the trajectories of private vehicles in the whole
Italy during May 2011.  This database refers, on average, to 2\% of
the vehicles registered in whole Italy, containing traces of
128,363,000 trips performed by 779,000 vehicles. Records are always
registered at engine starts and stops and every $\approx$2 km during
the trips (or alternatively every 30 seconds in the highways). Each
datum contains time, latitude-longitude coordinates, current velocity
and covered distance from the previous datum directly measured by
the GPS system using data recorded (but not registered) each second.
We define a trip as the transfer between two locations at which the
engine has been turned off. If the engine's downtime following a
stop is shorter than 30 seconds, the subsequent trajectory is
considered as a continuation of the same trip if it is not going
back towards the origin of the first trajectory. We have performed
filtering procedures to exclude from our analysis the data affected
by systematic errors ($\approx$ 10\% of data were discarded). The
problems due to signal loss is critical when the
engine is switched on or when the vehicle is parked inside a
building. In such cases we have used the information redundancy to
correct 20\% of the data by identifying the starting position of one
trip with the ending position of the previous one. When the signal
quality is good the average space precision is of the order of 10 m,
but in some cases it can reach values up to 30 meters or
more~\cite{bazzani2011}. Due to the Italian law on privacy, we have no direct
information on the owners or any specific knowledge about the social
characters of the drivers sample.

The GPS data base is collected for insurance reasons using black boxes installed on vehicles, whose owners
agreed with a special insurance contract. As a matter of
fact, these contracts are more attractive for young people or are used on fleet of vehicles.
This is a bias in our sample to study human mobility, since young people may use the private vehicle in a different way
with respect to elder people.  However our point of view is that the universal
statistical properties of human mobility discussed in the paper are not affected,
due to the large number trajectories and the different urban contexts. Some vehicles present in the database
belong to private companies' fleets. In this case, employers who use the car for professional reasons might show a different behaviour, but
they contribute to small percentage of all vehicles and therefore their statistical weight is small.

As the drivers's city of residence is unknown, it has been necessary
to associate each car to an urban area using the available
information. We have established that one driver lives in
a certain city if the most part of its parking time is spent in the
corresponding municipality area. For each driver, we have considered
all the mobility performed in a day (in and out the urban area) to
measure daily TTE $T$. In this way it is possible to measure the
average value of $T$ for over 1200 different municipalities, where
we have at least 100 vehicles. Moreover, for a smaller number of
cities we have sufficient data to analyse the shape of the
probability density $p(T)$ or of the cumulative distribution
$P(T)=\int_{0}^T p(T') dT'$, as
done in~\cite{gallotti2012} on a similar dataset.

\subsection*{A duration model for Travel-Time Expenditures}

\begin{figure*}
\includegraphics[angle=0, width=0.9\textwidth]{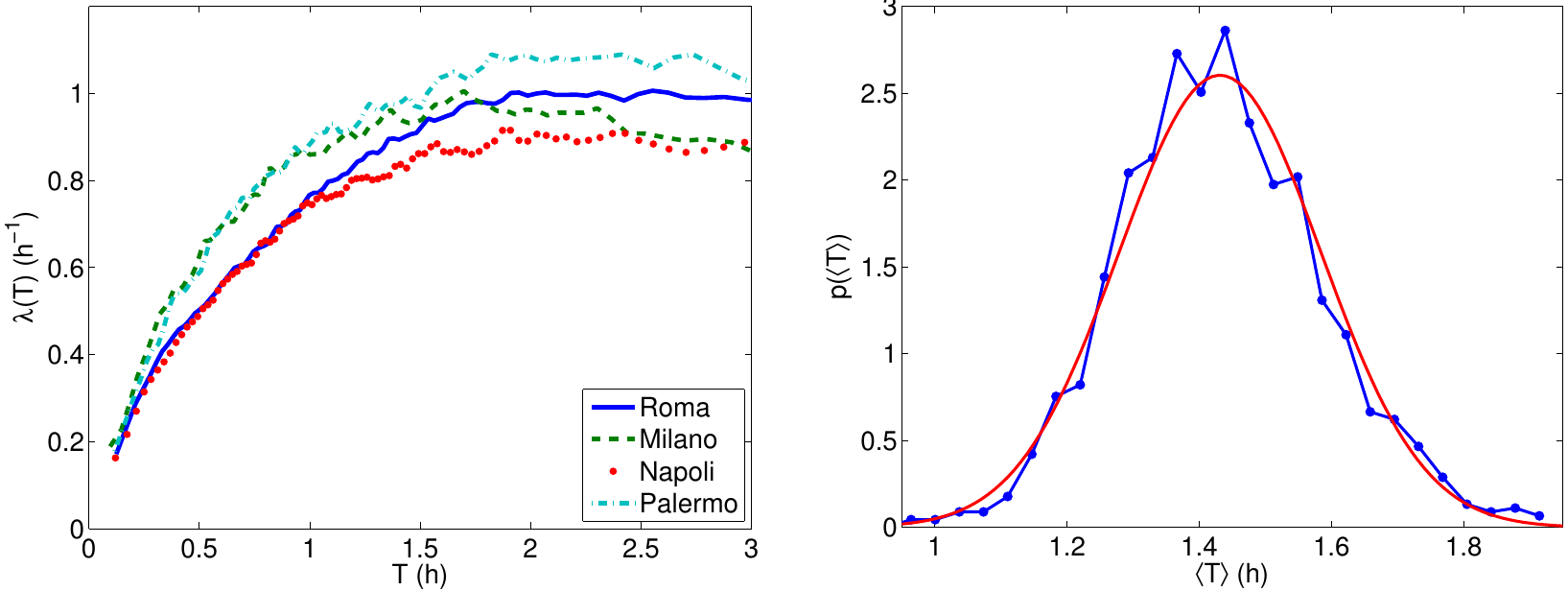}
\caption{ {\bf(Left) Different behaviours of the empirical hazard functions to model the mobility in a city.} Hazard functions $\lambda(T)$ for the cities
of Milan in Northern Italy with $1.3$ millions inhabitants (dash
line), Rome, the capital, in Central Italy with $2.7$  millions
inhabitants (solid line), Naples in Southern Italy $1$ million
inhabitants (dot line) and Palermo in Sicily $700$ thousand
inhabitants (dot-dash line). We have significant differences,
particularly in the height of the plateau (1/$\beta$) that is
related to the TTB. {\bf (Right) Average TTE in Italian cities.} The
distribution of the average TTE for 1233 Italian municipalities
where we have at least 100 GPS equipped vehicles (dots and lines)
can be interpolated with a Gaussian with mean 1.43h and standard
deviation 0.15h (solid line). } \label{differentTTE}
\end{figure*}

An application of duration model to
travel-time analysis has recently been proposed~\cite{juan2010}. This type of model
allows a mesoscopic description of the empirical data for a large
range of human and animal temporal behaviours~\cite{proekt2012}.

Using the GPS data base on single vehicle trajectories, it was possible to study the empirical TTE distribution
for all cities that had at least a sample of 100 monitored vehicles
(see Table S1). As example in Fig.~\ref{fitdistribution}, we show
the TTE empirical distributions for Naples, the largest cities in
the South of Italy ($\approx 1$ million inhabitants) and Grosseto, a
small city in the center of Italy ($\approx 80,000$ inhabitants).
This behaviour of the TTE distribution is observed in all
the considered cities. The parameter $\beta$, computed by
interpolating the empirical curves (see eq. (\ref{expdist})), defines
the average time scale of individual daily mobility and it is a
characteristic of each city. The distribution of the average TTE
$\langle T\rangle$ for those cities is reported in
Fig.~\ref{differentTTE} together with a normally distribution with
mean 1.43h, and standard deviation 0.15h. Those values are thus
significantly larger than the expected TTB of 1.1
hours~\cite{metz2008}.

From the comparison of definitions of TTB for different modes of
transportation, bodily energy consumption rates have to be taken
into account to define a universal travel-energy
budget~\cite{kolbl2003}. The TTB $\beta$ can be therefore
interpreted as a physiological limit to daily mobility: it is the
stress and fatigue accumulated while traveling that restricts the
time an individual is willing to spend on mobility in a day. Let $T$
the TTE of an individual, then we can introduce the \textit{survival
function} $S(T)$ as the probability that the TTE is greater than
$T$. Assuming the Markov properties for the evolution of $T$, we
have the relation
\begin{equation}
\label{defhT} S(T+ \Delta T) = [1 - \lambda(T)\Delta T]S(T)+o(\Delta
T)\,,
\end{equation}
where $\lambda(T)$ is the \textit{hazard function}, which is related
to the conditional probability $\pi(T+\Delta T|T)$ to realize a TTE
$T+\Delta T$ if one has spent a TTE $T$. The hazard function can be theoretically
defined as
\begin{equation}
\lambda(T)=\lim_{\Delta T\to 0} \frac{1-\pi(T+\Delta T|T)}{\Delta T} \,.
\end{equation}

If we consider an ensemble of individuals, the hazard function has
to be empirically defined as an average value
\begin{equation}
\lambda(T)=\left\langle\frac{1-\hat\pi(T+\Delta T|T)}{\Delta
T}\right \rangle_{\Delta T} \label{defhazard}
\end{equation}
where $\hat\pi(T+\Delta T|T)$ refers to the conditional probability
to observe a TTE $T+\Delta T$ of the individual dynamics and the
average value is computed over the distribution of the possible
increments $\Delta T$ in the considered population. $S(T)$ is
related to the probability distribution $p(T)$ with $p(T) =
-dS(T)/dT$. When the hazard function is constant, the underlying
stochastic process is a stationary Poisson distribution. But the
empirical hazard function, evaluated from GPS data (see
Fig.~\ref{differentTTE} left), shows an exponential decay from
the asymptotic uniform behaviour (see Fig. S1 and Supplementary
Information), which can be analytically interpolated by
\begin{equation}
\label{hazardFunction} \lambda(T) = \beta^{-1}[1-\exp(-T/\alpha)]
\,.
\end{equation}
We identify the parameter $\beta$ with the TTB, whereas $\alpha$ may
represent the typical average time associated the private car
mobility, since the hazard function $\lambda(T)$ is small when $T\le
\alpha$. As a matter of fact, both quantities are characteristic of
each city. The timescale $\alpha$ is associated to the accessibility
of desired destinations in the city~\cite{hansen1959}. Indeed, it is
interpreted as the average time necessary to satisfy the mobility
demand using private cars. Larger values of $\alpha$ mean lower
accessibility.
Given $\lambda(T)$, we can compute the
analytic form of the TTE probability distribution by explicitly
solving Eq. (\ref{defhT}) (see Supplementary Information)
\begin{equation}
\label{analyticaldistribution} p(T) =
\beta^{-1}\exp{(\alpha\beta^{-1})}\left(1-\exp(-T/\alpha)\right)\exp\left(-\alpha\beta^{-1}\exp(-T/\alpha)
- T/\beta\right) \,.
\end{equation}
According to Eq. (\ref{hazardFunction}), for $T \gg \alpha$ the
dominant term is $\exp(-T/\beta)$ and we recover the exponential
tail of the empirical TTE distributions. In the
Fig.~\ref{fitdistribution}, we show two interpolations of the
empirical distributions by using of the function
(\ref{analyticaldistribution}). The associated fits for the hazard
functions are displayed in Fig. S1. We have found a very good
agreement considering cities of different size, importance, position
and infrastructure development (see Table S1).

\subsection*{A time consumption model}
\label{stochasticModel}

To interpret the empirical results on an individual level, we
formulate a time consumption model where each individual
progressively accumulates travel-time according to a well defined
strategy. This interpretation is based on three key aspects:
\par i) the TTE is effectively a measure of the consumed \textit{Energy}~\cite{kolbl2003} during
mobility;
\par ii) there is a log-time perception of the trip durations as the TTE increases~\cite{helbing2010};
\par iii) the trip durations are exponentially distributed~\cite{gallotti2015}.
\par\noindent The first item refers to a Statistical Mechanics
interpretation of the TTE distribution function according to a
Maxwell-Boltzmann distribution. The second item means that after a
TTE of $T$, the perceived additional cost of a new trip by a driver
is proportional to $\Delta T/T$, where $\Delta T$ is the new
additional trip duration. The logarithmic scaling is a reflection of
Weber-Fechner psychophysical law~\cite{dehaene2003}. It is possible
that the individual perception of weariness is at the origin of this
logarithmic weighting of time, which has been proposed to explain
the statistical properties of the duration of individual
activities~\cite{gallotti2012}. The third item is supported by
empirical evidence: our data suggest that the travel-times cost $t$
for a single trip has also predominantly an exponential probability
density within the range $4 \leq t \leq 60$ minutes (see
Fig.\frenchspacing\ref{numerical} left)
\begin{equation}
p(t) \approx \langle t \rangle^{-1} \exp(-t/\langle t \rangle) \,,
\label{distritrip}
\end{equation}

\begin{figure*}
\includegraphics[angle=0, width=0.9\textwidth]{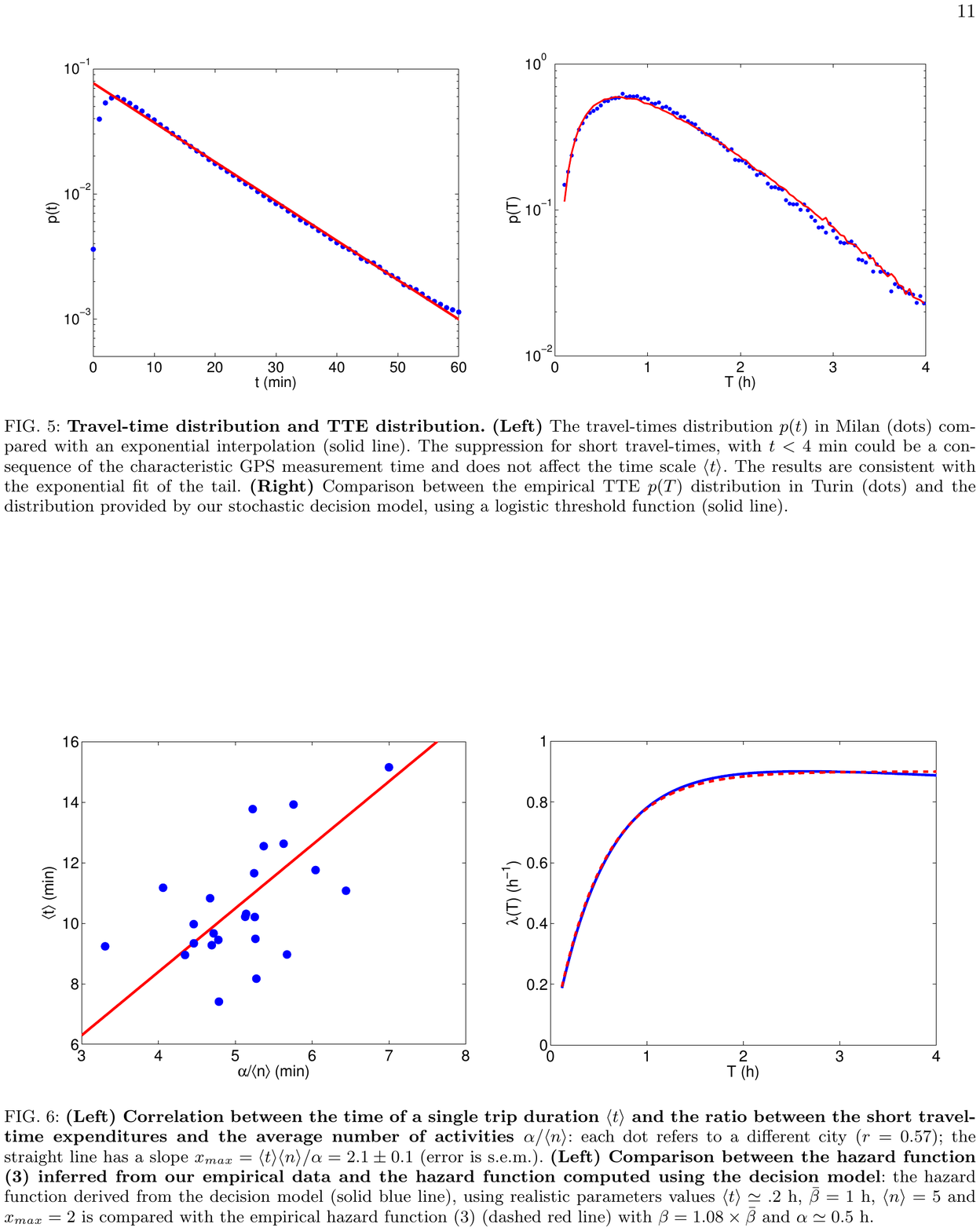}
\caption{\textbf{(Left) Correlation between the time of a single
trip duration $\langle t\rangle$ and the ratio between the short
travel-time expenditures and the average number of activities
$\alpha/\langle n\rangle$}: each dot refers to a different city ($r$
= 0.57); the straight line has a slope $x_{max}=\langle t\rangle
\langle n\rangle/\alpha=2.1\pm 0.1$ (error is s.e.m.).
\textbf{(Left) Comparison between the hazard function
(\ref{hazardFunction}) inferred from our empirical data and the
hazard function computed using the decision model}: the hazard
function derived from the decision model (solid blue line), using
realistic parameters values $\langle t\rangle \simeq .2$ h,
$\bar{\beta}=1$ h, $\langle n\rangle=5$ and $x_{max}=2$ is compared
with the empirical hazard function (\ref{hazardFunction}) (dashed
red line) with $\beta=1.08\times\bar\beta$ and $\alpha\simeq 0.5$
h.} \label{alpha}
\end{figure*}

This result has been shown to be universal across different cities, with the characteristic decaying time $\langle t \rangle)$ growing with city population~\cite{gallotti2015}.
In the Supplementary Information, we show that $\langle t\rangle$ also varies
among the considered cities and might depend upon house prices, city
surface and average travel speeds (See Fig. S2). In our model, each
individual progressively accumulates travel-time to determine his
TTE. According to our first assumption, a driver will accept a TTE
of $T$ with a probability
\begin{equation}
P(T)= \exp\left(-\frac{T}{\bar{\beta}}\right) \,, \label{time_ener}
\end{equation}
where $\bar{\beta}$ is the characteristic TTB of the population.
Then the individual conditioned probability to accept a new trip of
duration $\Delta T$ after a TTE of $T$ is written
\begin{equation}
\hat\pi(T+\Delta T/ T)= \frac{P(T+\Delta T)}{P(T)} \label{condprob}
\end{equation}
However the $\Delta T$ distribution for the new trip is not
independent from the elapsed TTE $T$ since users are reluctant to
accept long trips when the TTE exceeds $\bar{\beta}$. Then we define
a conditional $\Delta T$ distribution, which takes into account the
elapsed TTE, by using a threshold function $\theta_a(x)$
$$
\theta_a(x)=\begin{cases}  1 &\mbox{if } x<a \,, \\
 0 & \mbox{otherwise }. \end{cases}
$$
According to our assumptions, the distribution (\ref{distritrip}) is
substituted by the conditional distribution
\begin{equation}
p(\Delta T/T) \approx \langle t \rangle^{-1} \exp(-\Delta T/\langle
t \rangle)\theta_a(\Delta T/\gamma) \label{distritrip2}
\end{equation}
where the threshold $a$ and the time scale $\gamma$ depend on $T$ or
on other individual features: $a$ is the acceptability threshold for
a new trips, whereas $\gamma$ defines the perceived measure unit of
the cost of the new trip. The empirical observations suggest that
the threshold $a$ depends on the average number of activities
$\langle n\rangle$ of an individual. This is illustrated by the
correlation between the mobility timescale $\alpha$ (see eq.
(\ref{hazardFunction})) divided by $ \langle n\rangle$ and $\langle
t \rangle$ (see Fig.~\ref{alpha} left). To define $\gamma$, we
assume a logarithmic perception of the trip time cost so that
$\gamma\propto T$. Then we set the threshold $a=x_{max}/\langle n
\rangle$ and $\gamma=T$, so that the threshold function is
written in the form
$$
\theta_{x_{max}/\langle n \rangle}\left(\frac{\Delta T}{T}\right
)=\begin{cases}  1
&\mbox{if }  \frac{\Delta T}{T}< x_{max}/\langle n \rangle \,,\\
 0 & \mbox{otherwise }, \end{cases}
$$
where $x_{max}$ turns out to be an \textit{universal threshold}.
Therefore Eq. (\ref{distritrip2}) is based on the assumption that
the propensity of a driver to accept a further trip of duration
$\Delta T$ after having performed a TTE of $T$, scales as $\langle
n\rangle/T$, where $\langle n\rangle$ is the average number of daily
activities. In other words the individuals that perform more trips
using private vehicles have a greater TTB: this could be also a
consequence of the the multi-modal mobility, which is not included
in our database, and that allows individuals to divide their TTB
according the different transportation means used. Moreover, an
individual seems to organise the mobility using the TTB as a
mobility energy (with the constraint of performing the compulsory
daily activities), but keeping the percentage of TTE fluctuations
constant. Using empirical values for the different quantities in the
relation (\ref{alfaapp}) we can estimate $x_{max}\simeq 2$ (see
Fig.~\ref{alpha} left).
\par\noindent
We compute the empirical hazard function for a population of drivers
according to the definition (\ref{defhazard})
$$
\lambda(T)=\langle t \rangle^{-1} \int_0^\infty \frac{1-\exp(-\Delta
T/\beta)}{\Delta T}\theta_{x_{max}/\langle n
\rangle}\left(\frac{\Delta T}{T}\right) \exp(-\Delta T/\langle t
\rangle)d\Delta T
$$
An explicit calculation (see Supplementary Information) shows that
the hazard function of the model has the same analytic form as the
empirical interpolation (\ref{hazardFunction}), where the timescale
of the short TTE suppression is
\begin{equation}
\alpha\simeq \langle n\rangle \langle t\rangle /x_{max} \,,
\label{alfaapp}
\end{equation}
as one can see in Fig.~\ref{alpha} (right).

\section*{Competing interests}
The authors declare that they have no competing interests.

\section*{Author's contributions}
RG, AB designed and performed research and wrote the paper; RG, AB, SR performed research. RG prepared the figures.

\section*{Acknowledgements}
We thank D. Helbing for useful comments on an early draft. We thanks Octo Telematics S.p.A. for providing the GPS database. RG thanks M. Barthelemy, G. Carra, Y. Crozet, M. Lenormand, T. Louail and R. Louf for useful discussions at the QuantUrb seminars.

\clearpage
\section{Supplementary Information}

\renewcommand*{\thefigure}{S\arabic{figure}}
\renewcommand*{\thetable}{S\arabic{table}}

\newcounter{defcounter}
\setcounter{defcounter}{0}

\newenvironment{SIequation}{
\addtocounter{equation}{-1}
\refstepcounter{defcounter}
\renewcommand\theequation{S\thedefcounter}
\begin{equation}}
{\end{equation}}

\setcounter{figure}{0}

\subsection{List of analyzed cities}

\begin{table*}[h!]
\centering
\ra{1.3}
 \begin{tabular*}{\textwidth}{@{\extracolsep{\fill}}rlrrccccccccc@{}}
  \toprule[1pt]
 \# & Name & Pop. & Area & v & $\langle t\rangle$ & $\langle n\rangle$ & $\langle T\rangle$ &$\alpha$ & $\beta$ & $R^2$ \\
\hline
 1 &            Roma & 2786034 &  1285.3  & 26.87 & 0.253 & 5.87  & 1.52 & 0.69$\pm$0.01 & 1.00$\pm$0.01 & 0.99\\
 2 &         Palermo &  653522 &   158.9 & 23.00 & 0.170 & 6.80  & 1.39 & 0.60$\pm$0.01 & 0.92$\pm$0.01 & 0.99\\
 3 &          Genova &  606978 &   243.6 & 23.17 & 0.211 & 5.34 & 1.29 & 0.50$\pm$0.02  & 0.88$\pm$0.02 & 0.98\\
 4 &          Napoli &  957012 &   117.3  & 24.47 & 0.232 & 6.35  & 1.58 & 0.61$\pm$0.01 & 1.11$\pm$0.01 & 0.99\\
 5 &            Bari &  340355 &   116.0 & 31.03 & 0.172 & 6.77 & 1.52 & 0.58$\pm$0.02  & 1.04$\pm$0.02 & 0.99\\
 6 &          Milano & 1345890 &   181.8 & 24.35 & 0.230 & 5.52  & 1.44  & 0.48$\pm$0.01& 1.02$\pm$0.01 & 0.99\\
 7 &          Torino &  905780 &   130.3  & 24.17 & 0.196 & 5.68 & 1.36 & 0.57$\pm$0.02 &  0.88$\pm$0.01 & 0.97\\
 8 &         Bologna &  383949 &   140.7  & 29.22 & 0.180 & 5.94 & 1.37 & 0.46$\pm$0.02 & 0.95$\pm$0.01 & 0.97\\
 9 &         Firenze &  373446 &   102.4  & 25.93 & 0.209 & 5.76 & 1.40 & 0.52$\pm$0.02  & 0.98$\pm$0.02 & 0.95\\
10 & Reggio Calabria &  186273 &   236.0 & 24.79 & 0.123 & 8.32 & 1.42 & 0.66$\pm$0.02 & 0.90$\pm$0.01 & 0.99\\
11 &         Perugia &  169290 &   449.9  & 35.22 & 0.149 & 7.05 & 1.36  & 0.67$\pm$0.04& 0.83$\pm$0.02 & 0.98\\
12 &         Catania &  289971 &   180.9  & 23.95 & 0.170 & 7.12 & 1.48 & 0.61$\pm$0.02  & 0.99$\pm$0.01 & 0.99\\
13 &          Foggia &  152181 &   507.8  & 24.20 & 0.149 & 7.61 & 1.45  & 0.55$\pm$0.03 & 0.98$\pm$0.02 & 0.99\\
14 &           Forl\`i &   118968 &   228.2  & 33.30 & 0.186 & 6.61 & 1.27 & 0.45$\pm$0.02  & 0.87$\pm$0.02 & 0.99\\
15 &        Grosseto &  82616 &   474.3  & 31.80 & 0.154 & 6.98 & 1.19  & 0.39$\pm$0.02 & 0.83$\pm$0.02 & 0.99\\
16 &          Latina &  120526 &   277.8  & 31.32 & 0.156 & 7.19 & 1.40 & 0.53$\pm$0.03 & 0.95$\pm$0.02 & 0.99\\
17 &           Lecce &  96274 &   238.4  & 31.46 & 0.155 & 8.33 & 1.46 & 0.65$\pm$0.04  & 0.94$\pm$0.02 & 0.98\\
18 &         Messina &   240858 &   211.2  & 23.13 & 0.185 & 6.89 & 1.43  & 0.74$\pm$0.03 & 0.89$\pm$0.02 & 0.98\\
19 &          Modena &   186289 &   183.2  & 32.09 & 0.158 & 6.65 & 1.30 & 0.53$\pm$0.03  & 0.83$\pm$0.02 & 0.97\\
20 &           Parma &  189833 &   260.8  & 33.87 & 0.166 & 6.45 & 1.37 & 0.48$\pm$0.02 & 0.93$\pm$0.02 & 0.98\\
21 &         Ravenna &   159404 &   244.2 & 38.18 & 0.161 & 6.58 & 1.31 & 0.52$\pm$0.03  & 0.86$\pm$0.02 & 0.98\\
22 &   Reggio Emilia &  172317 &   231.6  & 32.00 & 0.158 & 6.48 & 1.29& 0.57$\pm$0.03  & 0.81$\pm$0.02 & 0.97\\
23 &         Salerno &  138284 &    59.0  & 28.16 & 0.194 & 6.55 & 1.49 & 0.57$\pm$0.03 & 1.03$\pm$0.02 & 0.99\\
24 &        Siracusa &   123517 &   204.1 & 24.43 & 0.136 & 7.46 & 1.30 & 0.66$\pm$0.03 & 0.82$\pm$0.02 & 0.99\\

\bottomrule[1pt]
 \end{tabular*}
 \caption{The different columns are: \# = id number in the figures,
Name = municipality where the driver spent most of his parking time,
Pop.
 = Municipality Population, Area = Municipality Area (km$^2)$, v = speed (km/h), $\langle t\rangle$ = average travel-time per trip (h), $\langle n\rangle$ = average number of daily trips, $\langle T\rangle$ = average daily travel time expenditure (h), $\alpha$  =  accessibility measure (h), $\beta$ = travel-time budget (h), $R^2 =$ coefficient of determination.  Errors represent the 95\% confidence intervals for the fit the parameters, estimated using a bootstrap method with 100 repetitions.}

\end{table*}

\clearpage

\subsection{Duration Model}

\begin{figure}[h!]
\centerline{
\includegraphics[angle=0, width=0.45\textwidth]{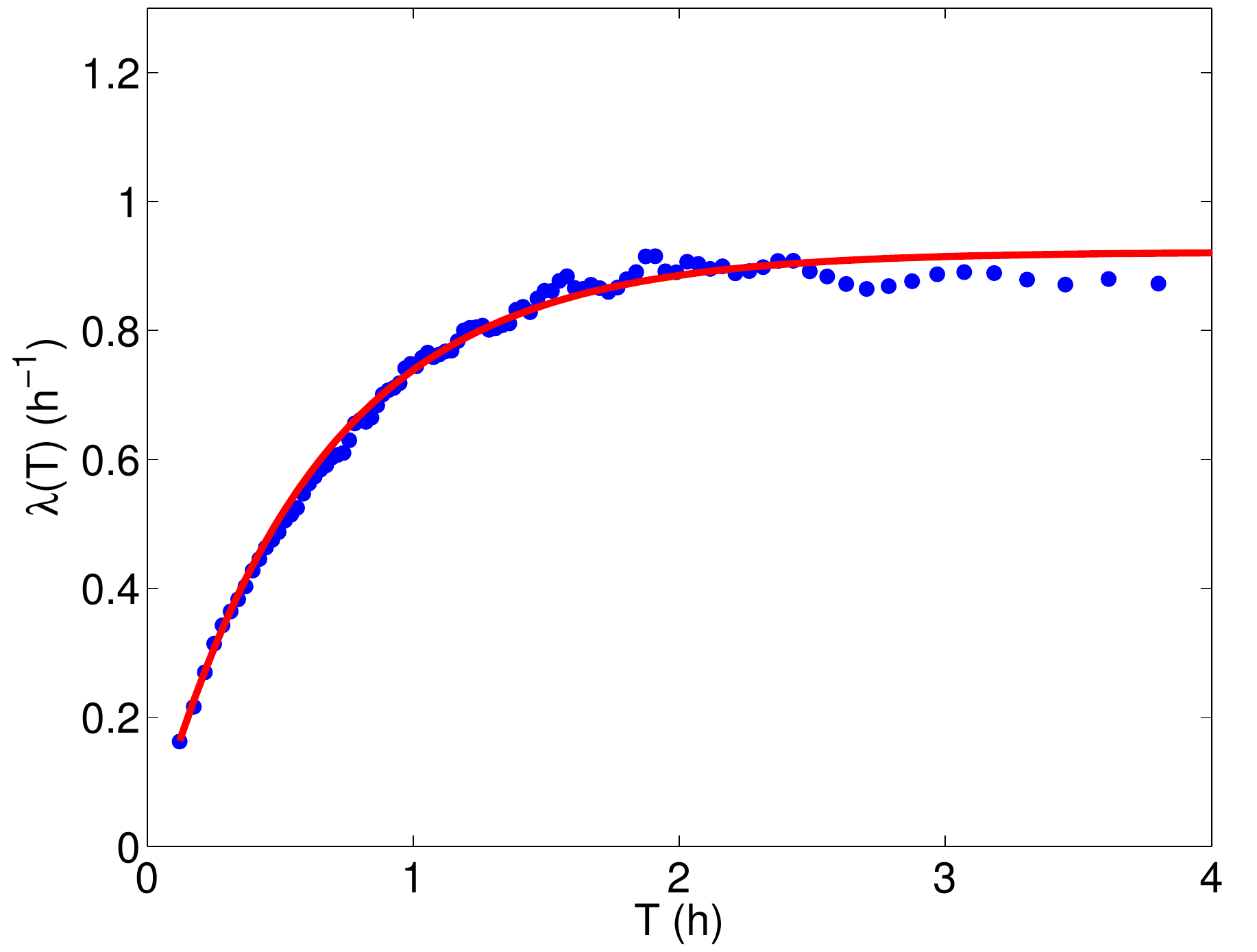}
\quad
\includegraphics[angle=0, width=0.45\textwidth]{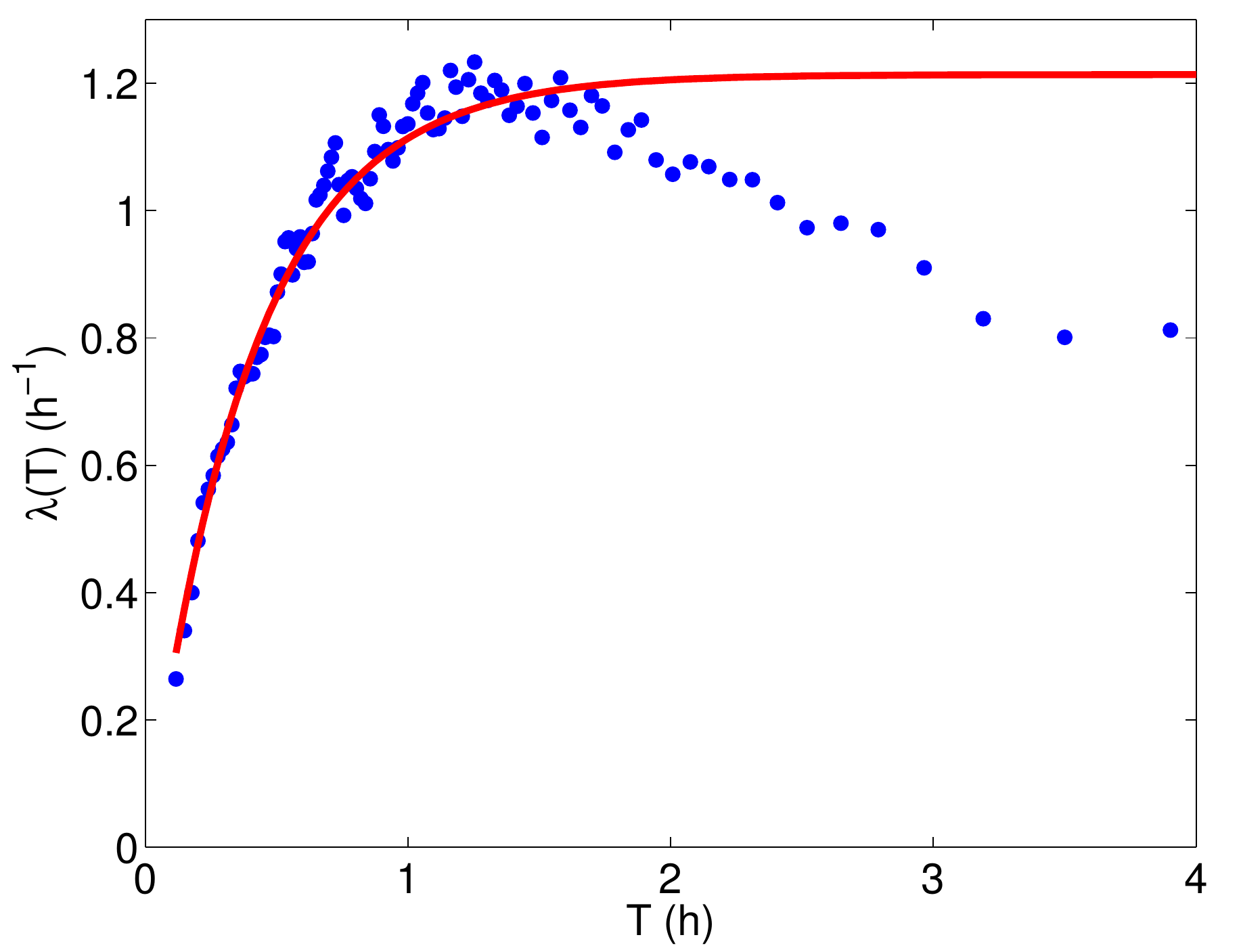}
} \vspace*{8pt}
\caption{\textbf{(Left) Hazard function in Naples.}
The empirical hazard function $\lambda(T)$ (dots) is found to
be exponentially converging to a constant value. \textbf{(Right)
Hazard function in Grosseto.}
We remark that the lack of data in the TTE distribution tail does not allow an accurate numerical evaluation of the hazard function. Therefore, in this case to estimate of $\alpha$
and $\beta$ in some cities, we exclude the decreasing behaviour of the empirical hazard function. In both figures, the
values of the parameters obtained by the exponential fit of
$\lambda(T)$(solid line) define the solid line in Fig.~1.}
\label{hazardfunctionFigSI}
\end{figure}
The proposed TTE model is based on the Markov property for the
evolution of the survival distribution $S(T)$ (see def.
(2) in the main paper). Assuming a regular character for the decision
stochastic process, we have
\begin{SIequation}
S(T+\Delta T)=\pi(T+\Delta T| T) S(T)+o(\Delta T) \,,
\label{markov}
\end{SIequation}
where the likelihood $\pi(T+\Delta T|T)$ defines the conditioned
probability for the representative individual of performing a daily
mobility $T+\Delta T$ given the elapsed time $T$ from the beginning
of the daily mobility. Then, we introduce the \textit{hazard
function} $\lambda(T)$ (i.e. the probability of ending the mobility after
a time the interval $\Delta T$ knowing that the user was still driving at time $T$) according to
\begin{SIequation}
\lambda(T)=\lim_{\Delta T\to 0}\frac{1-\pi(T+\Delta T| T)}{\Delta
T}\label{defhazard} \,,
\end{SIequation}
However the definition (\ref{defhazard}) cannot be used if we
consider a microscopic stochastic dynamics that mimics the
individual decisions, so in the paper , we propose an alternative
definition which link the individual behaviour with the average
property of the population
\begin{SIequation}
\lambda(T)=\left\langle\frac{1-\hat\pi(T+\Delta T/T)}{\Delta
T}\right \rangle_{\Delta T} \label{SIdefhazard2}
\end{SIequation}
where $\hat\pi(T+\Delta T/T)$ is the conditional probability to
observe a TTE $T+\Delta T$ of the individual dynamics and the
average value is computed over the distribution of the possible
increments $\Delta T$ in the considered population.
\par\noindent
In the limit $\Delta T\to 0$ from (\ref{markov}) we get the
differential equation:
\begin{SIequation}
\label{diffeq} dS(T)/dT = -\lambda(T)S(T) \,.
\end{SIequation}
In a stationary situation, the hazard function $\lambda(T)$ is constant
($\lambda(T) = h_0$) and the differential equation (\ref{diffeq}) leads to
an exponential solution $S(T) = h_0 \exp(-h_0 T)$ which corresponds
to an exponential probability density ($p(T) = -dS(T)/dT$) for the
TTE. Under this point of view the exponential tail of the empirical
distribution $p(T)$ can be associated to a constant probability of
ending the daily mobility, independently from the elapsed daily
travel-time $T$ as expected for a Poisson process. However, the
observed under-expressed short travel-times implies an increasing
trend for the hazard function $\lambda(T)$: i.e. when the TTE is short,
the probability to stop the daily mobility is lower than the
asymptotic value. This observation suggests that it is unlikely for an individual to consider his daily
mobility concluded after a very short cumulative travel-time $T$,
because some daily duties have still to be accomplished. A more
suitable shape of the function $\lambda(T)$ can be extrapolated from GPS
data (see Supplementary Fig.~\ref{hazardfunctionFigSI}). We perform
an analytical interpolation of the empirical data by
\begin{SIequation}
\label{hazardFunctionSI} \lambda(T) =-\frac{dS(T)/dT}{S(T)} =
\beta^{-1}\left(1-\exp(-T/\alpha)\right) \,,
\end{SIequation}
where the parameters $\beta$ is the TTB, characteristic of a
particular city, and $\alpha$ is the timescale of the short travel
time expenditures suppression.  According to our point of view
$\alpha$ could be interpreted as the average time necessary to
satisfy the daily mobility demand using private cars. After a time
$\alpha$, the choice of going back home is only due to the limited
TTB constraint, quantified by the time scale $\beta$. Given $\lambda(T)$,
we integrate analytically the differential equation (\ref{diffeq})
obtaining an analytical form for the survival function
\begin{SIequation}
S(T) = C_N \exp\left(-\alpha\beta^{-1}\exp(-T/\alpha) -
T/\beta\right) \,,
\end{SIequation}
Imposing $S(0) = 1$ (thus neglecting null TTEs), the normalisation constant can be fixed at $C_N = \exp{(\alpha\beta^{-1})}$. Consequently, the probability density function for the TTE distribution reads
\begin{SIequation}
\label{analyticaldistributionSI} p(T) =
\beta^{-1}\exp{(\alpha\beta^{-1})}\left(1-\exp(-T/\alpha)\right)\exp\left(-\alpha\beta^{-1}\exp(-T/\alpha)
- T/\beta\right) \,.
\end{SIequation}

\subsection{Analytical solution for the time consumption model}

Let us consider a driver who has carried out a daily mobility $T$
and he has to decide if to perform or not a further trip whose
duration is $\Delta T$. According to our Statistical Mechanics point
of view, the exponential decay of the empirical TTE distribution (see eq. (1) and Fig.~1 in the paper), suggests that the mobility time
plays the role of the \textit{energy}. As a consequence we expect
that the probability to accept a TTE $T$ is
\begin{SIequation}
P(T)=\exp\left(-\frac{T}{\bar{\beta}}\right ) \,,
\end{SIequation}
where $\bar{\beta}$ is the expected value of TTE. In the model, to
evaluate the probability of performing a new trip the drivers
considers the possibility to accept the cost $\Delta T$ of the new
trip using a threshold function
\begin{SIequation}
\theta_{x_{max}/\langle n \rangle}\left(\frac{\Delta T}{T}\right
)=\begin{cases}  1
&\mbox{if }  \frac{\Delta T}{T}< x_{max}/\langle n \rangle \,,\\
 0 & \mbox{otherwise }, \end{cases}
\label{logtime}
\end{SIequation}
where $\langle n\rangle$ is the average number of performed daily
activities and $x_{max}$ is an universal threshold (see Fig.~5
left). Then according to empirical observations (see. eq. (9) in the
text) we introduce the conditional distribution for the time cost
$\Delta T$ of the single trips
\begin{equation}
p(\Delta T/T) \approx \langle t \rangle^{-1} \exp(-\Delta T/\langle
t \rangle)\theta_{x_{max}/\langle n \rangle}\left(\frac{\Delta
T}{T}\right ) \label{SIdistritrip2}
\end{equation}
The presence of the threshold function $\theta_{x_{max}/\langle
n\rangle}$ means that, as the TTE increases, the individual gets
used to be driving and he has a propensity to accept longer trips
(compatibly with his TTB and the number of activities he has to
perform). Since the perceived cost of a new trip in the model is set
$\propto \Delta T/T$, we correlate this choice with the existence of
a \textit{log-time} perception. To reproduce the macroscopic
statistical laws of human mobility, the drivers are considered as
\textit{independent particles} and we average on the cost $\Delta T$
of the individual trips using the empirical distribution
(\ref{SIdistritrip2}) (Supplementary Fig.~\ref{singletriptimes}
(a)). According to the TTB existence assumption, a rational driver
evaluates the probability to perform the new trip after having used
a TTE $T$, as
\begin{SIequation}
\pi(T+\Delta T|T)= \frac{P(T+\Delta
T)}{P(T)}=\exp\left(-\frac{\Delta T}{\bar{\beta}}\right ) \,.
\end{SIequation}
and using the definition (\ref{SIdefhazard2}) of the hazard
function, we set
\begin{SIequation}
\lambda(T)=\left \langle \frac{1-\pi(T+\Delta T|T)}{\Delta T}\right
\rangle=\int_0^\infty \left (\frac{1-\exp(-\Delta
T/\bar{\beta})}{\Delta T}\right )\theta_{x_{max}/\langle
n\rangle}\left(\frac{\Delta T}{T}\right ) d\Delta T \,,
\label{inthazard}
\end{SIequation}
and using the definition (\ref{logtime}) we derive an analytical
expression for the hazard function
\begin{SIequation}
\lambda(T)=\frac{1}{\alpha}\int_0^{x_{max}}\exp\left
(-\frac{Tx}{\langle n\rangle \alpha}\right
)\frac{1-\exp(-Tx/(\langle n\rangle \bar\beta))}{x}dx \,,
\label{approx}
\end{SIequation}
where we introduce the timescale $\alpha$ (see eq.
(\ref{hazardFunctionSI}))
\begin{SIequation}
\alpha\simeq \langle n\rangle \langle t\rangle /x_{max} \,.
\label{alphath}
\end{SIequation}
The time scale (\ref{alphath}) is consistent with the empirically evaluated timescale for the short TTE suppression (see Fig.~5
left) with $x_{max}\simeq 2$. A numerical integration of eq.
(\ref{approx}) provides a hazard function which has the same
behaviour as the interpolation (\ref{hazardFunctionSI}) derived from
the empirical GPS data. In the Fig.~5 right, we show a comparison
between the integral (\ref{approx}) and the empirical hazard
function where $\beta=1.08 \bar\beta$ and $\alpha$ computed from the
previous relation. We remark the presence of a scaling factor
between the empirical evaluated $\beta$ and the theoretical expected
value $\bar \beta$. More precisely $\beta$ proves to be an
overestimate of $\bar \beta$ since the incremental ratio in the
integral (\ref{inthazard}) decreases as the cost $\Delta T$ becomes
large. In other words, according to the time consumption model, the
empirical data bestow a greater TTB to individuals with respect to
the theoretical value, due to the reduced perception of the trip
cost when the TTE increases.

\subsection{Properties of the average trips' duration $\langle t \rangle$ and average number of trips $\langle n\rangle$.}

\begin{figure}[h!]
\begin{center}
\begin{tabular}{cc}
\raisebox{2.3cm}{(a)} \includegraphics[angle=0, width=0.45\textwidth]{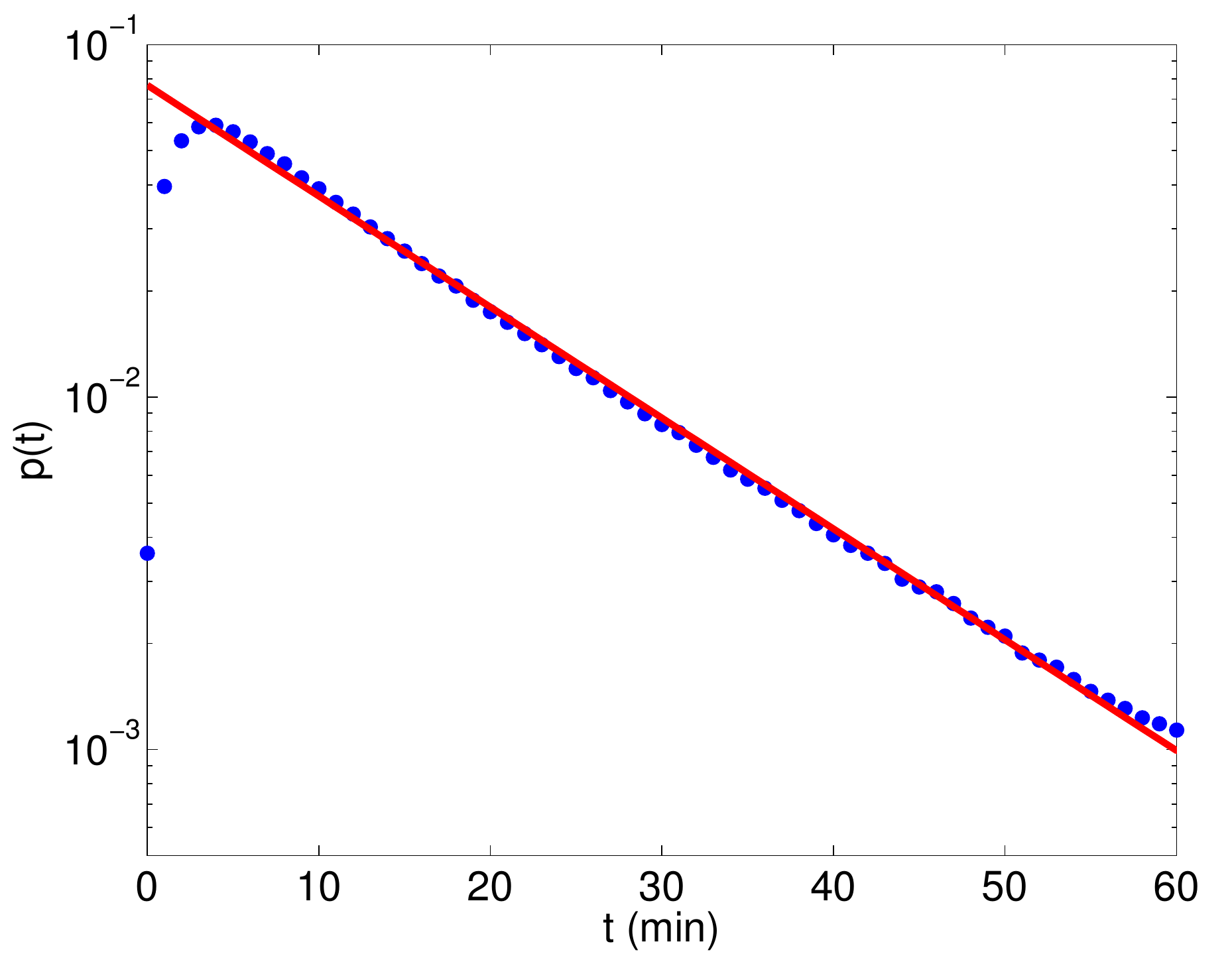}&
\raisebox{2.3cm}{(b)} \includegraphics[angle=0, width=0.45\textwidth]{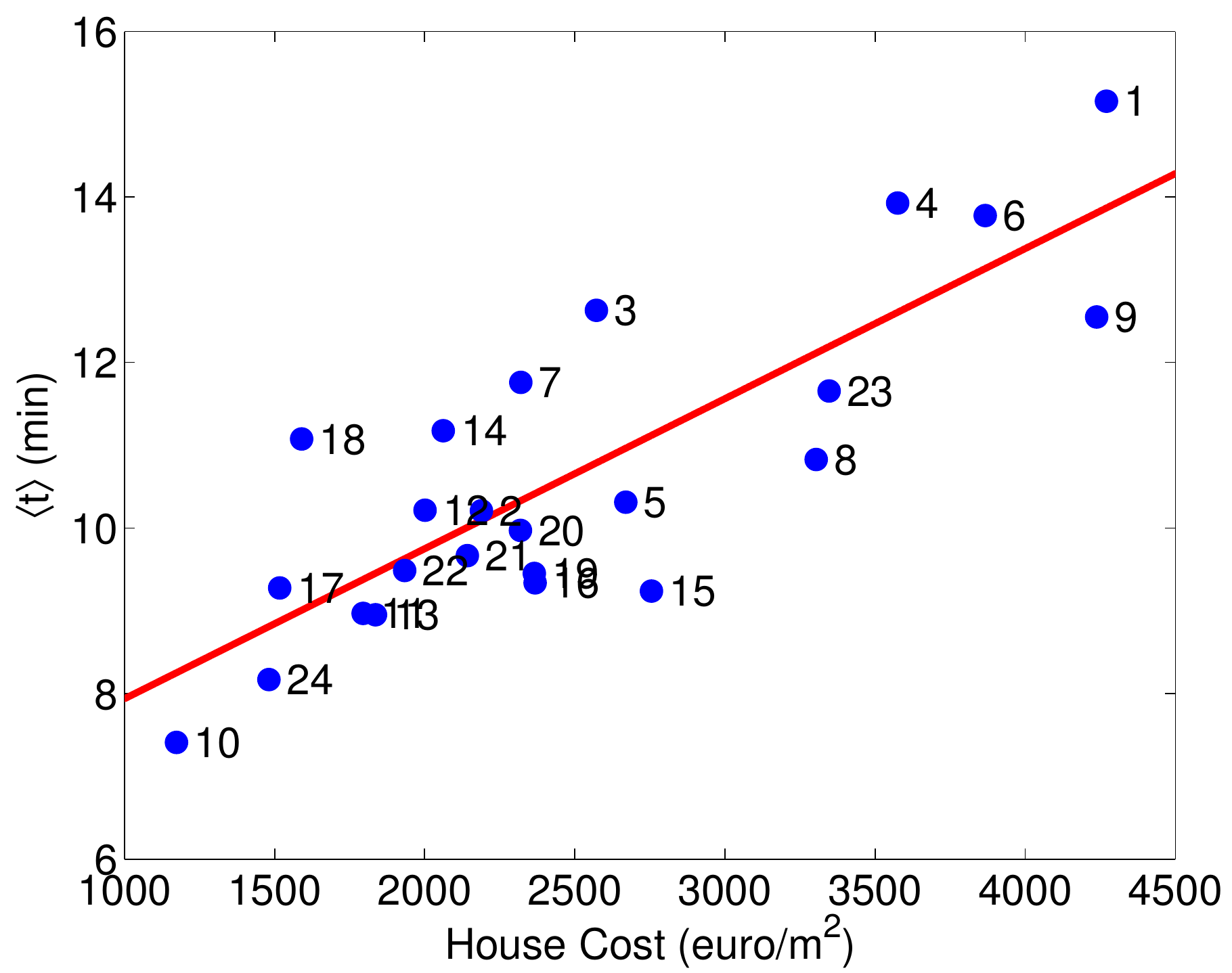} \\
\raisebox{2.3cm}{(c)} \includegraphics[angle=0, width=0.45\textwidth]{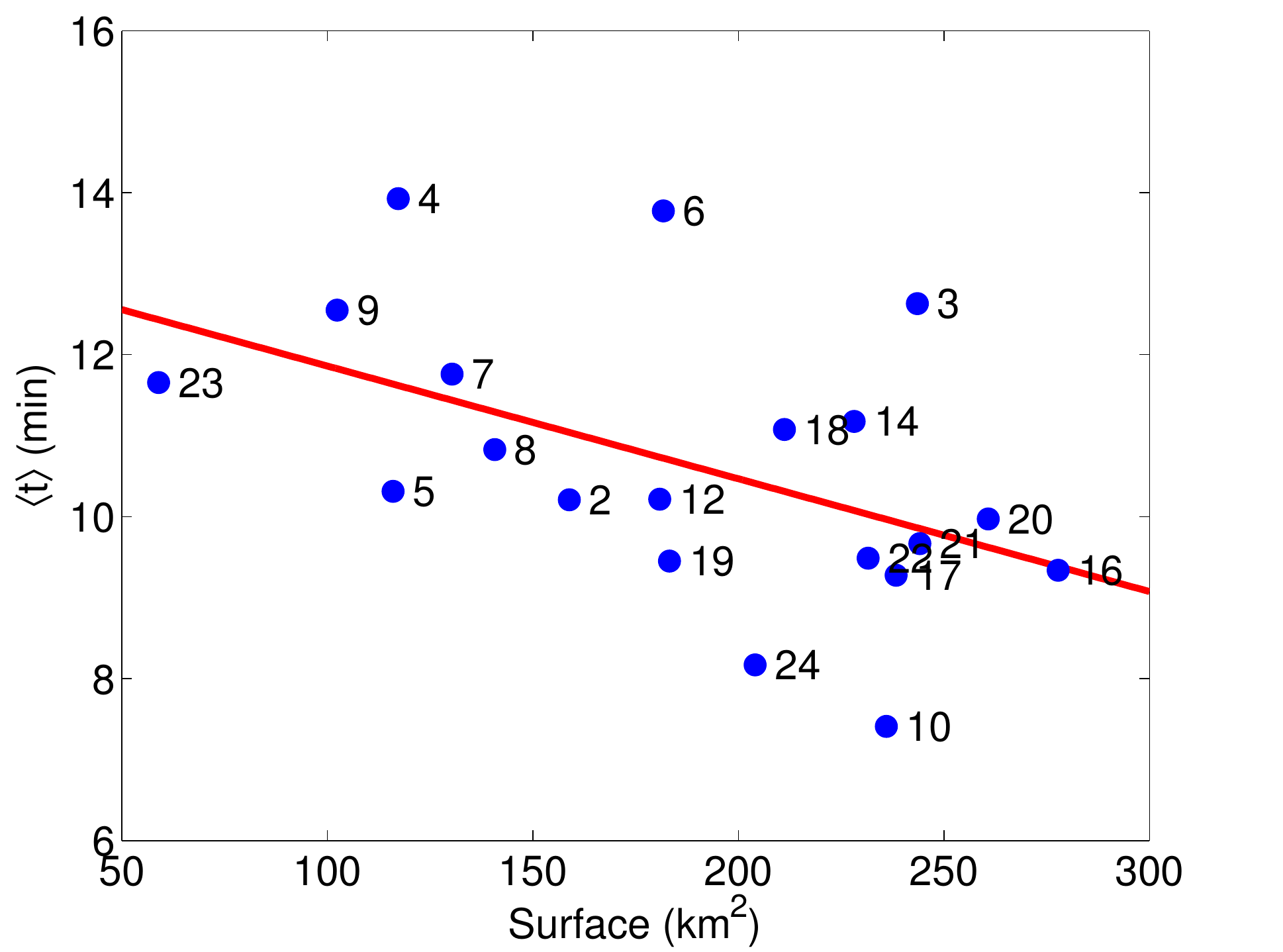} &
\raisebox{2.3cm}{(d)} \includegraphics[angle=0, width=0.45\textwidth]{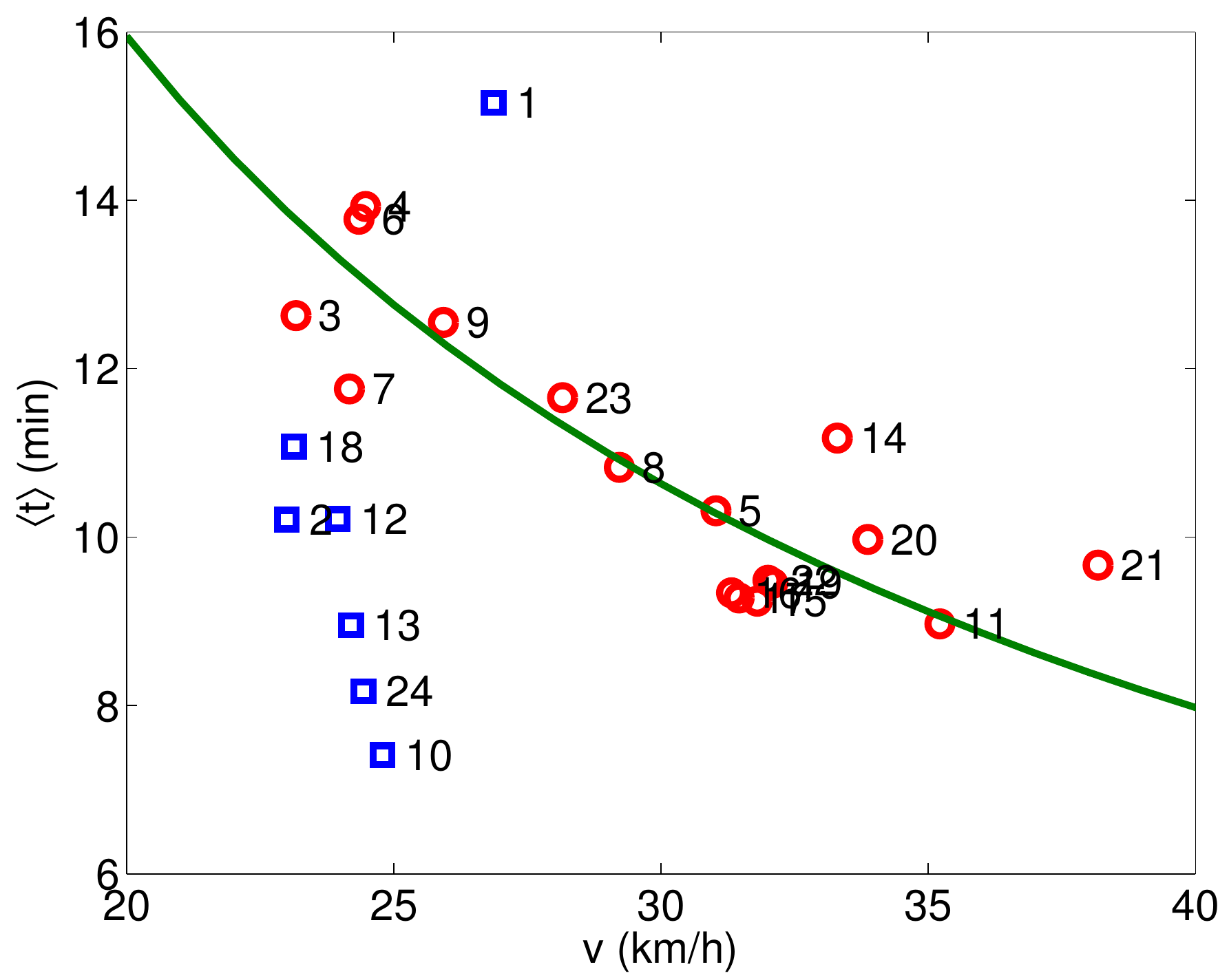} \\
\end{tabular}
\end{center}
\caption{{\bf Properties of the average trips' duration $\langle t \rangle$} \textbf{(a) Exponential distribution.} The $p(t)$
distribution (dots) for Milan and exponential interpolation of its tail
(solid line) with $\langle t \rangle = 13.8$min;  \textbf{(b)
Growth with house prices}.
Travel-times grow significantly in cities where housing is more
expensive (correlation coefficient 0.83, source: www.immobiliare.com); \textbf{(c) Decrease with the
city surface.}  Travel-times tend
to be reduces in ider cities (correlation coefficient -0.49);
\textbf{(d) Decrease with travel speed.}  A part of the cities lie approximatively on an
hyperbole (solid line), representing a constant average length of
$\approx 5.3 km$.} \label{singletriptimes}
\end{figure}

We consider the correlation between the parameters $\alpha$
and $\beta$ with the average travel-time $\langle t\rangle$ for a
single trip. The results show that $\langle t\rangle$ has a
positive correlation of 0.57 with $\beta$ and no correlation with
$\alpha$. But $\langle t\rangle$ is
strongly correlated with the average house costs per square
meter (Supplementary Fig.~\ref{singletriptimes} (b)) and negatively correlated with
the municipalities surface (Supplementary Fig.~3 (b)). This
empirical observation could be a consequence of the activities
sprawling in the larger cities, whereas they are concentrated inside
the historical center for the smaller cities. The relationship between
$\langle t\rangle$ and the average trip's speed seems instead not
trivial (Supplementary Fig.~(d)). As a matter of fact an
almost constant average trip length of $\approx 5.3$ km is observed
in the majority the cities, independently by the municipality area.
Therefore one expect a relation $\langle t\rangle v=const$ among the
cities, where $v$ is the average travel speed characteristic of the different
 road networks. Indeed, if we exclude Rome, whose spatial
scale is much larger than that of all the other cities, the cities
with an average speed greater than $25$ km/h verify this relation,
whereas we observe a strong deviation from the theoretical curve in
the cities with average speed $<25$ km/h. We interpret this effect
as the result of a different dynamic regime in the road network:
when the average travel speed is low the stochastic effects due to
the stops at crossings or to congestion effects could strongly
influence the vehicle dynamics, so that the proportionality between
covered distance and time is lost. On the contrary a high average
travel speed suggests that the free flow is dominant and the vehicle
dynamics can be described in a deterministic way.

If one computes the number of daily trips $n$, whose empirical
distribution $p(n)$ shows an exponential tail~\cite{gallotti2012}, we
see that the limiting average values of $\langle n\rangle$ are
strongly anti-correlated (correlation coefficient -0.78) with the
average trip length $\langle t \rangle$, suggesting a tradeoff
consistent with the concept of TTB. This seems confirmed by the
negligible correlation (correlation coefficient -0.17) between
$\langle n\rangle$ and $\beta$, whereas between $\langle n\rangle$
and $\alpha$ the correlation is 0.40, reflecting the role of
$\alpha$ as a measure of the time needed for the necessary mobility.
Finally, we have a remarkably low correlation (correlation
coefficient 0.13) between the average number of daily trips $\langle
n\rangle$ and the average trip's speed, which confirms that mobility
induced by travel-time savings is not due to a larger number of
trips but to longer trips~\cite{metz2008}.

\clearpage
\subsection{Disaggregated analysis for the city of Milan}

To study the effect of individual heterogeneity we have
disaggregated the empirical data into different classes of drivers.
This analysis has been performed for the city of Milan. Due to the
absence of any metadata the features to characterize the individuals
have been extracted from the GPS data according to:
\begin{itemize}
\item home location, identified by the parking place where the cumulative parking time is the longest one~\cite{bazzani2010};
\item number of days in which the individual have used the car during the month;
\item structure of the mobility network: mono-centric or polycentric~\cite{gallotti2013}.
\end{itemize}

To identify differences in TTE dependence from home location we have
divided the Milan municipality in three concentric areas, according
to the central structure of the city. We have chosen circular
boundaries that we can approximatively associate with:
\begin{itemize}
\item[i)] the area within the inner ring road (Cerchia dei Bastioni)
identified as the \textit{Zona C}, the name that identifies the
congestion charge area;
\item[ii)] the area between the inner and the outer ring
road (Cerchia dei Navigli), that we call \textit{city center} in
Fig. 3 (b);
\item[iii)] the \textit{periphery}, outside the outer ring road.
\end{itemize}
Among the drivers identified as citizens of Milan, 7\% live in the
Zona C, 27\%  live in the city center and 63\% live in the
periphery. The remaining 3\% are individuals whose home locations we
found outside the city area and they were excluded from the
analysis. To point out differences in the home's role, we take into
account all the mobility performed in and out the municipality area
of Milan, evaluating the percentage $rt$ of round trips involving
home as origin or destination. When $rt>75\%$ we define the
individual mobility network as mono-centric: 58\% of the drivers in
Milan have this property. Conversely, if $rt<75\%$, we can introduce
a second hub in the individual mobility network~\cite{gallotti2013},
which has a significative role in the organization  of the
individual mobility.

\subsection{Numerical formulation of the time consumption model}

Each individual accumulates progressively the travel times into the total travel-time $T_n = \sum_{i=1}^n t_i$, where $n$ is the number of daily trips. From the other hand each trip is associated to a performed activity and it is possible to introduce an utility function
$U$~\cite{helbing2010}, representing in some preference scale the satisfaction and/or the advantages derived by performing
that activity. Without any further information, our null hypothesis is that the activity utility $U$ is a random variable
uniformly distributed in the interval $(0,1)$ (arbitrary units). To define a behavioural model we assume that
this utility is counter-balanced by a cost due to the time already spent driving until that moment.

Each trip represents an increment of total travel time $\Delta T = t$. Travel-times $t$ are distributed exponentially with average fixed at the experimental values of table S1.
The total cost associated to travel is not to be quantified not proportionally to $T$, but to its logarithm $\log(T)$
plus a certain constant to exclude negative values.  Then the
cumulative utility $\mathcal{U}$ is given by the linear combination:
\begin{equation}
\mathcal{U} = c_1(c_2 U - \log(T)) \label{util}
\end{equation}
The parameter $c_2$ represents the cost/benefit ratio, which could be associated to a value of time), while
$c_1$ is the unit measure of the utility scale which is associated to the shape of the logistic threshold. $T$ is the individual
TTE distributed according to eq. (\ref{analyticaldistributionSI}).

If we evaluate the probability of performing a daily activity
according to the logistic model~\cite{domencich1975}
\begin{equation}
p(U) = \exp(\mathcal{U})/(1+\exp(\mathcal{U})) \label{logistic}
\end{equation}
using Monte-Carlo simulations, the related TTE distribution turns
out to be very similar to the empirical one (see Fig.~3 right). For all considered cities, the best fits have a $R^2>0.986$. Similarly to what observed in the main text, this correspondence does not happen assuming costs proportional to $T$.

\end{document}